\documentclass[conference]{IEEEtran}
\IEEEoverridecommandlockouts
\usepackage{cite}
\usepackage{amsmath,amssymb,amsfonts}
\usepackage{algorithmic}
\usepackage{graphicx}
\usepackage{textcomp}
\usepackage{placeins}
\usepackage{xcolor}
\usepackage{booktabs}
\usepackage{amssymb}
\usepackage{amsthm}

\usepackage[hyphens]{url}
\usepackage{xurl}
\usepackage[table]{xcolor}
\definecolor{rowgray}{gray}{0.93}
\definecolor{ourred}{RGB}{170,30,30}
\def\BibTeX{{\rm B\kern-.05em{\sc i\kern-.025em b}\kern-.08em
    T\kern-.1667em\lower.7ex\hbox{E}\kern-.125emX}}

\begin{document}

\title{Map2APS: A Physically Grounded Benchmark for Direct Angle Power Spectrum Prediction from Urban Geometry
}

\author{\IEEEauthorblockN{Junxi Huang,
Xiucheng Wang,
Nan Cheng,
Kailong Wang,
Ruijin Sun,
Zhisheng Yin
}

\IEEEauthorblockA{
\IEEEauthorrefmark{1}State Key Laboratory of ISN and School of Telecommunications Engineering, Xidian University, Xi'an, 710071, China\\
Email: \{24012100067, xcwang\_1, 24012100069\}@stu.xidian.edu.cn, \\dr.nan.cheng@ieee.org, \{sunruijin, zsyin\}@xidian.edu.cn}
}

\maketitle

\begin{abstract}
Angle power spectrum (APS) characterizes the directional distribution of received signal power and is directly relevant to beam management and MIMO processing. While environment-aware learning has been widely studied for radio maps and path loss, direct map-to-APS prediction still lacks a standardized large-scale benchmark. This paper presents Map2APS, a physically grounded benchmark constructed from intelligent ray-tracing (IRT) path-level propagation records. Map2APS covers 51 equal-height urban maps and approximately 2.55 million Tx--Rx samples, with a strict cross-map split for evaluating generalization to unseen urban layouts. We benchmark representative model families and introduce MS-AReg as a strong reference baseline. On the full held-out test set of 249{,}993 samples, MS-AReg achieves a cosine similarity of 0.948, a peak location error of 1.20$^\circ$, and an inference latency of 0.101 ms/sample. We further report dominant-direction metrics, including Top-1 dominant peak hit rate and dominant peak recall, to evaluate whether predicted spectra preserve decision-relevant arrival directions. The benchmark, code, and evaluation scripts are released at \url{https://github.com/UNIC-Lab/aps-data}.
\end{abstract}

\begin{IEEEkeywords}
angle power spectrum, benchmark dataset, map-to-APS prediction, 
environment-aware learning, dominant arrival direction
\end{IEEEkeywords}

\section{Introduction}

The angle power spectrum (APS) characterizes the directional distribution of received signal power and is directly relevant to beam management and MIMO processing \cite{rappaport2013mmwave,heath2016mmwave}. Conventional APS acquisition typically relies on ray tracing or large-scale field measurements, which are computationally expensive when repeated over many transmitter--receiver links and urban layouts \cite{3gpp38901}. Learning-based environment-aware wireless prediction has been widely studied for radio-map estimation and path-loss prediction~\cite{alkhateeb2019deepmimo,levie2021radiounet,zhang2023rmegan,wang2025radiodiff}, and more recently extended to differentiable ray tracing, channel knowledge maps, and learning-based radio map construction~\cite{hoydis2023sionnart,wang2026tutorial,wu2024ckmbeam}. However, direct map-to-APS prediction remains largely unexplored, and no standardized benchmark exists for comparing model families, evaluating cross-map generalization, or assessing whether predicted spectra preserve decision-relevant arrival directions.

APS prediction differs from radio-map or path-loss prediction because its output is a structured directional spectrum rather than a scalar power value or a spatial coverage field. The quality of an APS prediction depends not only on point-wise amplitude accuracy, but also on global spectral-shape consistency and accurate localization of dominant angular peaks. In urban scenarios, local geometric structures such as building boundaries, street openings, and corner regions may shift dominant arrival directions even when average received power remains similar. Therefore, a map-to-APS benchmark should evaluate spectral-shape agreement, peak-localization accuracy, and dominant-direction preservation in addition to conventional reconstruction errors.

This paper presents \textbf{Map2APS}, a physically grounded benchmark for direct APS prediction from urban geometry. Given a rasterized urban map and transmitter/receiver locations, the task is to predict a 180-bin APS over the azimuthal arrival-angle range without running ray tracing at inference. APS labels are constructed from intelligent ray-tracing (IRT) path-level propagation records \cite{altair_irt}, and the benchmark contains 51 equal-height urban maps with approximately 2.55 million Tx--Rx samples under a strict cross-map split. We further benchmark representative model families under a unified protocol and introduce MS-AReg as a strong reference baseline.

The main contributions are summarized as follows:
\begin{enumerate}
    \item We introduce \textbf{Map2APS}, a standardized benchmark for direct APS prediction from urban geometry, focusing on structured directional spectra rather than scalar power maps or path-loss values.
    \item We construct a physically grounded large-scale dataset from IRT path-level propagation records, with compact 180-bin APS labels and a strict cross-map split for evaluating generalization to unseen urban layouts.
    \item We provide a representative benchmark suite and a unified evaluation protocol covering spectral-shape, peak-localization, latency, and dominant-direction metrics, with MS-AReg serving as a strong reference baseline.
\end{enumerate}

\section{Map2APS Benchmark Construction}

We consider an urban propagation scenario in which all buildings have the same height. Propagation is modeled in the horizontal plane. For each Tx--Rx link, the input is a three-channel condition image composed of a binary rasterized building map and two Gaussian heatmaps centered at the transmitter and receiver:
\begin{equation}
\mathbf{c}
=
\left[
M_{\mathrm{bldg}};
G(\mathbf{p}_{\mathrm{tx}},\sigma);
G(\mathbf{p}_{\mathrm{rx}},\sigma)
\right]
\in \mathbb{R}^{3\times H\times W}.
\end{equation}
Here, $M_{\mathrm{bldg}}$ is the building map, and $G(\mathbf{p},\sigma)$ is a Gaussian heatmap centered at $\mathbf{p}$ with standard deviation $\sigma$. The three channels encode the building geometry and the Tx/Rx positions.

For each Tx--Rx pair, an intelligent ray-tracing (IRT) simulator provides path-level propagation records \cite{altair_irt}. Each record contains the delay, arrival angle, and received power of one propagation path. Let the path set of a link be
\begin{equation}
\mathcal{P}
=
\{(\tau_k,\theta_k,p_k)\}_{k=1}^{K},
\end{equation}
where $\tau_k$, $\theta_k$, and $p_k$ denote the delay, arrival angle, and linear-domain received power of the $k$-th path, respectively. To construct the APS label, path contributions are aggregated using a delay-domain interpolation kernel and an angular array-response kernel:
\begin{equation}
\begin{gathered}
Q(\tau,\theta)
=
\sum_{k=1}^{K}
p_k\,
\mathrm{sinc}^2\!\bigl(f_s(\tau-\tau_k)\bigr)
\left[
\frac{\sin(N\psi_k)}
     {N\sin(\psi_k)}
\right]^2,\\
\psi_k
=
\pi d_\lambda(\sin\theta-\sin\theta_k).
\end{gathered}
\end{equation}
Here, $f_s$ is the bandwidth-related sampling rate, $N$ is the number of antenna elements, and $d_\lambda$ is the antenna spacing normalized by wavelength. The function $Q(\tau,\theta)$ denotes the path-aggregation quantity used for APS label construction.

The final APS label is obtained by retaining the dominant directional energy at each angular bin:
\begin{equation}
a(\theta_j)
=
\max_{\tau} Q(\tau,\theta_j),
\qquad
j=1,\ldots,180,
\end{equation}
where $\{\theta_j\}_{j=1}^{180}$ denotes the discretized azimuthal angle grid over $[-180^\circ,180^\circ)$ with $2^\circ$ resolution. This operation yields a compact directional power spectrum that is aligned with the objective of direct map-to-APS prediction.

After converting the raw APS to the dB domain, we normalize it and convert it back to the linear power domain for learning. Specifically, the dB-domain APS is first shifted by its sample-wise maximum, converted to the linear power domain, and standardized using dataset-level statistics:
\begin{equation}
\tilde{\mathbf{a}}_{\mathrm{dB}}
=
\mathbf{a}_{\mathrm{dB}}
-
\max_j a_{\mathrm{dB},j},
\qquad
\mathbf{a}_{\mathrm{lin}}
=
10^{\tilde{\mathbf{a}}_{\mathrm{dB}}/10},
\end{equation}
\begin{equation}
\bar{\mathbf{a}}
=
\frac{\mathbf{a}_{\mathrm{lin}}-\mu_a}{s_a}.
\end{equation}
This preprocessing removes sample-wise absolute-scale variation and makes the learning target focus on relative directional spectrum shape and dominant peak structure.

Given the benchmark dataset
\begin{equation}
\mathcal{D}
=
\{(\mathbf{c}^{(n)},\bar{\mathbf{a}}^{(n)})\}_{n=1}^{N_s},
\end{equation}
the task is to learn a predictor
\begin{equation}
F_{\Theta}:\mathbb{R}^{3\times H\times W}\rightarrow\mathbb{R}^{180}
\end{equation}
that maps the condition image to the normalized APS. The benchmark learning objective is
\begin{equation}
\min_{\Theta}
\frac{1}{N_s}
\sum_{n=1}^{N_s}
\mathcal{L}
\left(
F_{\Theta}(\mathbf{c}^{(n)}),
\bar{\mathbf{a}}^{(n)}
\right).
\end{equation}
This formulation is model-agnostic: different regression architectures can be evaluated under the same input representation, APS label construction rule, data split, and metric protocol.

\begin{figure*}[!t]
    \centering
    \includegraphics[width=0.98\textwidth]{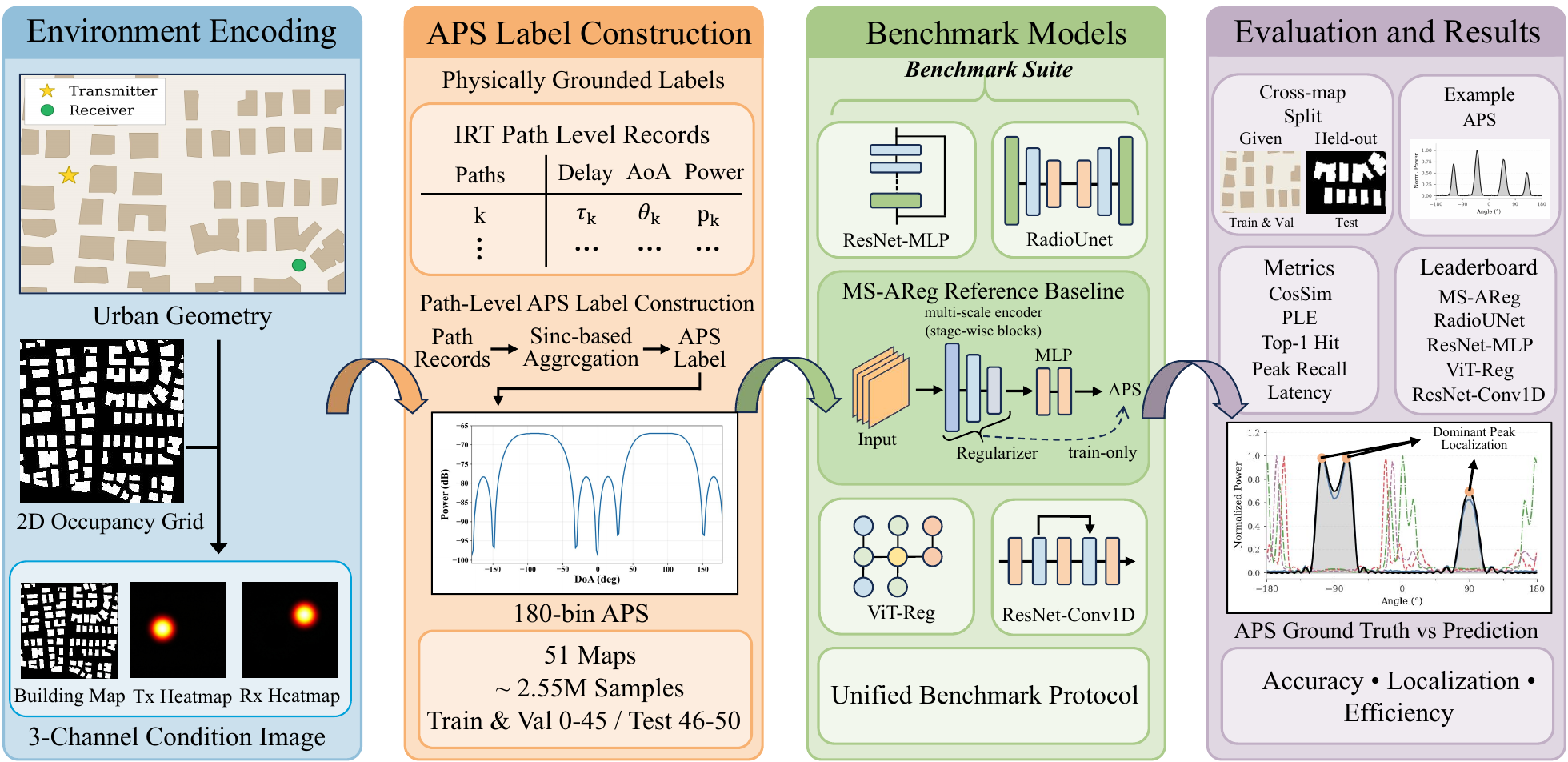}
    \caption{Map2APS benchmark construction pipeline. The input condition image is formed from the urban building map and Tx/Rx heatmaps. APS labels are constructed from IRT path-level propagation records through sinc-based path aggregation and APS label generation. The benchmark uses map-level train/test separation and a unified evaluation protocol covering spectral-shape, peak-localization, latency, and dominant-direction metrics.}
    \label{fig:Structure}
\end{figure*}

The benchmark dataset contains 51 equal-height urban maps generated under the same IRT settings, with 100 transmitter locations and 500 receiver locations per map, yielding approximately 2.55 million samples in total. Maps 0--45 are used for training and model development, while maps 46--50 are reserved as a held-out test set with no map overlap, providing a strict evaluation of cross-map generalization. The benchmark construction pipeline is illustrated in Fig.~\ref{fig:Structure}.

\section{Benchmark Models and Reference Baseline}

To evaluate Map2APS under a unified protocol, we include representative regression models with different architectural biases. ResNet-MLP represents single-scale CNN regression with global pooling, RadioUNet represents UNet-style convolutional modeling \cite{levie2021radiounet}, ViT-Reg represents transformer-style global token modeling \cite{dosovitskiy2021vit}, and ResNet-Conv1D provides an additional multi-scale regression comparison with a convolutional spectral decoder. In addition, we introduce MS-AReg as a strong reference baseline for this benchmark. Together, these models cover the main architectural design axes relevant to map-to-APS prediction: spatial encoding strategy, decoder type, and model scale.

\subsection{MS-AReg Overview}

MS-AReg maps the three-channel condition image $\mathbf{c}$ to the normalized APS $\hat{\mathbf{a}}\in\mathbb{R}^{180}$ in a deterministic regression manner. It consists of a multi-scale ResNet-18 encoder, an MLP predictor, and a training-only regularizer. The encoder extracts multi-level geometric features from the input condition image, the predictor decodes the fused representation into the APS vector, and the regularizer provides additional condition-spectrum structural supervision during training. The regularizer is discarded at inference, so the deployed model only contains the encoder and predictor.

\subsection{Multi-Scale Encoder}

Direct map-to-APS prediction is sensitive to geometric structures at multiple spatial scales. Dominant APS peaks depend not only on the global Tx--Rx layout, but also on local building boundaries, street openings, and corner regions. A single deepest feature after global pooling may capture scene-level semantics, but it can suppress localization-critical spatial details. A multi-scale encoder that aggregates features across spatial scales is therefore better aligned with the structure of this task. MS-AReg extracts feature maps from four stages of a pre-trained ResNet-18 backbone \cite{he2016resnet}.

Let $F_i=E_i(\mathbf{c})$ denote the feature map at stage $i$. Each stage is adaptively pooled and linearly projected as
\begin{equation}
\begin{gathered}
\mathbf{s}_i
=
W_i \cdot
\mathrm{flatten}\!\left(
\mathrm{AdaptiveAvgPool}(F_i)
\right)
+b_i \in \mathbb{R}^{256},\\
i=1,2,3,4 .
\end{gathered}
\end{equation}
Specifically, the four stages are pooled to $4\times4$, $2\times2$, $1\times1$, and $1\times1$ spatial resolutions, respectively. The projected vectors are then concatenated and fused into a global condition representation:
\begin{equation}
\mathbf{g}
=
\mathrm{MLP}
\left(
[\mathbf{s}_1;\mathbf{s}_2;\mathbf{s}_3;\mathbf{s}_4]
\right)
\in
\mathbb{R}^{512}.
\end{equation}
This representation combines boundary-level details, intermediate spatial topology, and global layout context, making it suitable for dominant-peak localization and spectral-shape prediction.

\begin{figure*}[!t]
    \centering
    \includegraphics[width=\textwidth]{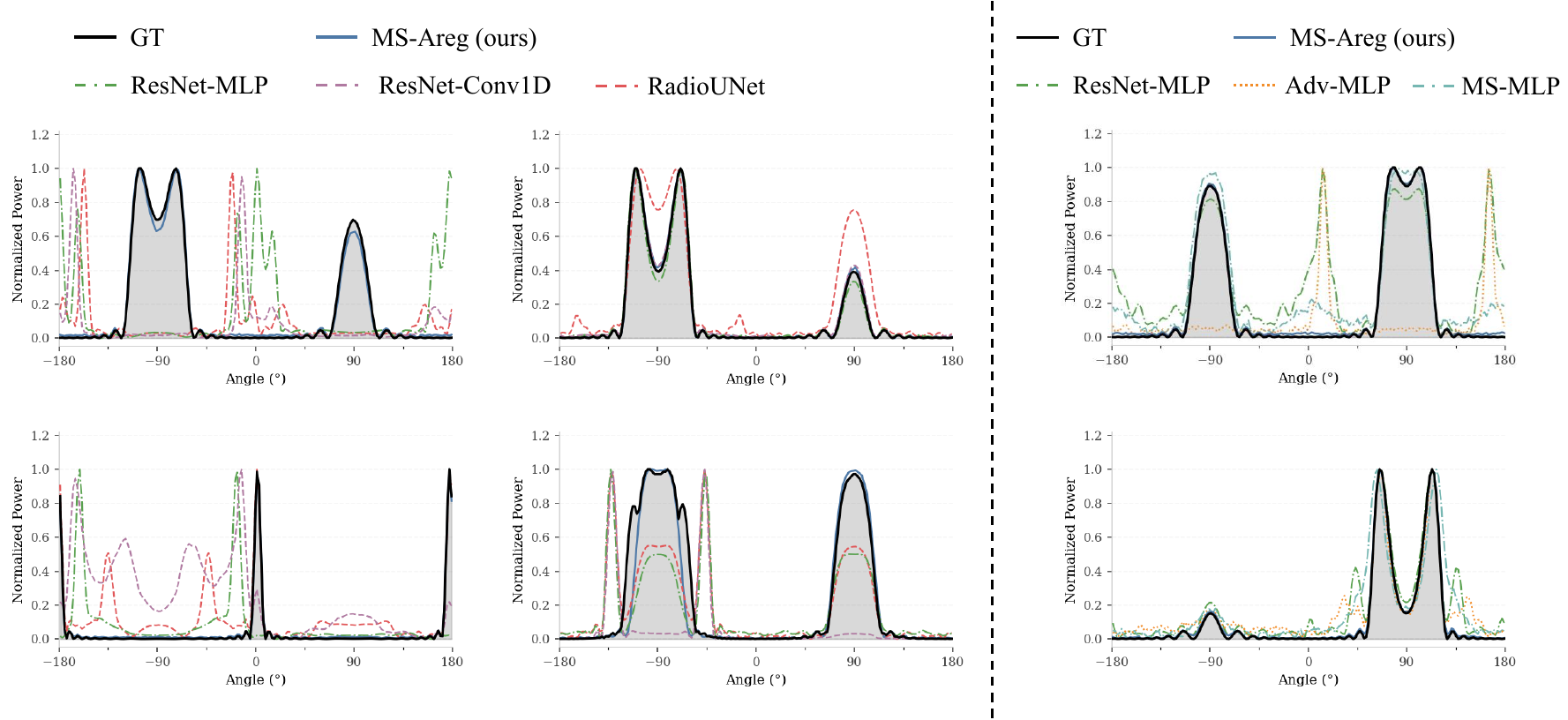}
    \caption{Representative qualitative comparison of APS prediction on held-out test samples. The left side forms the main-comparison zone, where the ground-truth APS is overlaid with predictions from MS-AReg, RadioUNet, ResNet-MLP, and ResNet-Conv1D. The right side forms the ablation zone, where the ground truth is compared against MS-AReg, ResNet-MLP, Adv-MLP, and MS-MLP under the same overlay-style visualization. Across both zones, MS-AReg more accurately aligns dominant peaks and better preserves the overall spectral profile.}
    \label{fig:qualitative}
\end{figure*}

\subsection{MLP Predictor}

The fused condition representation $\mathbf{g}$ is decoded by a four-layer MLP into a 180-dimensional APS vector. The decoder contains three hidden layers with widths 1024, 1024, and 512, followed by a 180-dimensional output layer, with LeakyReLU activations between adjacent hidden layers. Since the target APS is a global directional spectrum rather than a spatial output map, the MLP decoder is sufficient for deterministic spectrum regression while keeping the inference pipeline lightweight.

\subsection{Training-Only Regularization}

Point-wise losses such as L1 or MSE constrain angular bins largely independently and may produce overly smoothed APS estimates. To provide additional structural supervision, MS-AReg uses a training-only regularizer that scores the compatibility between the condition image and an APS. Given a pair $(\mathbf{c},\mathbf{a})$, where $\mathbf{a}$ denotes either a normalized real APS $\bar{\mathbf{a}}$ or a predicted APS $\hat{\mathbf{a}}$, the regularizer outputs a scalar score $R(\mathbf{c},\mathbf{a})$. In practice, the regularizer is implemented with separate encoders for the condition image and the APS, followed by a scoring head.

The regularizer and predictor are optimized with
\begin{equation}
\mathcal{L}_{R}
=
\mathrm{BCE}\!\left(R(\mathbf{c},\bar{\mathbf{a}}),1\right)
+
\mathrm{BCE}\!\left(R(\mathbf{c},\hat{\mathbf{a}}),0\right),
\end{equation}
\begin{equation}
\mathcal{L}_{P}
=
\mathrm{BCE}\!\left(R(\mathbf{c},\hat{\mathbf{a}}),1\right)
+
\lambda_{L1}
\|\hat{\mathbf{a}}-\bar{\mathbf{a}}\|_1 .
\end{equation}
The L1 term anchors point-wise reconstruction accuracy, while the regularization term encourages spectra that are structurally compatible with the input geometry. During regularizer updates, $\hat{\mathbf{a}}$ is detached from the predictor to keep the optimization roles separated. At inference, the regularizer is removed completely, and MS-AReg remains a pure deterministic regressor.

\section{Experiments and Benchmark Results}

\subsection{Experimental Setup}

The experimental data follow the Map2APS benchmark protocol described in Sec.~II. Maps 0--45 are used for training and model development, while maps 46--50 are reserved as the held-out test set with no map overlap. All quantitative results are reported on the full held-out test set of 249{,}993 samples.

Using the same dominant-peak selection rule as in PLE evaluation,
75\% of the held-out samples contain at most two dominant peaks,
with an average of 2.28 dominant peaks per sample. This highlights
the sparse but localization-sensitive structure of the Map2APS labels.

All models are trained and evaluated under the same normalized APS target domain and the same cross-map split. For MS-AReg, both the predictor and the regularizer are optimized using Adam with a learning rate of $2\times10^{-4}$ and $(\beta_1,\beta_2)=(0.5,0.999)$. The model is trained for 20 epochs with batch size 1024 in bfloat16 mixed precision. The loss weight is set to $\lambda_{L1}=100$. Inference latency is measured as the average per-sample runtime on the full held-out test set.

\begin{table*}[t]
\centering
\caption{Main benchmark results on the full held-out test set of 249{,}993 samples. Best results are shown in bold.}
\label{tab:main_results}
\small
\setlength{\tabcolsep}{5pt}
\begin{tabular}{lccccccc}
\toprule
Method & MAE$\downarrow$ & PSNR$\uparrow$ & CosSim$\uparrow$ & NMSE$\downarrow$ & Angle$\downarrow$ & PLE$\downarrow$ & ms/sample$\downarrow$ \\
\midrule
\rowcolor{rowgray}
\textcolor{ourred}{\textbf{MS-AReg}} &
\textcolor{ourred}{\textbf{0.099}} &
\textcolor{ourred}{\textbf{37.23}} &
\textcolor{ourred}{\textbf{0.948}} &
\textcolor{ourred}{\textbf{-20.03}} &
\textcolor{ourred}{\textbf{10.08$^\circ$}} &
\textcolor{ourred}{\textbf{1.20$^\circ$}} &
0.101 \\
RadioUNet & 0.182 & 34.11 & 0.861 & -16.91 & 19.10$^\circ$ & 1.30$^\circ$ & 0.933 \\
ResNet-MLP & 0.193 & 31.51 & 0.851 & -14.31 & 21.06$^\circ$ & 1.69$^\circ$ & \textbf{0.088} \\
ViT-Reg & 0.216 & 33.70 & 0.804 & -16.50 & 24.00$^\circ$ & 2.06$^\circ$ & 0.822 \\
ResNet-Conv1D & 0.224 & 31.36 & 0.802 & -14.16 & 25.18$^\circ$ & 2.29$^\circ$ & 0.091 \\
\bottomrule
\end{tabular}
\end{table*}

\subsection{Evaluation Metrics}

We evaluate all methods using MAE, PSNR, NMSE, cosine similarity (CosSim), spectral angle, peak location error (PLE), dominant-direction metrics, and inference latency. MAE, PSNR, and NMSE measure point-wise reconstruction accuracy, while CosSim and spectral angle measure global spectral-shape agreement. Lower MAE, NMSE, spectral angle, PLE, and latency indicate better performance, whereas higher PSNR, CosSim, Hit Rate, and Recall are preferred.

PLE is computed using a dynamic dominant-peak selection rule. For each ground-truth APS, local peaks are detected after max normalization with a prominence threshold of 0.05. Let the detected peak heights be $\{h_i\}$. Their relative contributions are
\begin{equation}
r_i=\frac{h_i}{\sum_j h_j+\epsilon},
\end{equation}
and peaks with $r_i\ge 0.1$ are retained as dominant peaks. If no peak satisfies this condition, the highest peak is retained. PLE is defined as
\begin{equation}
\mathrm{PLE}
=
\frac{1}{|\mathcal{P}_{gt}|}
\sum_{\theta_i\in\mathcal{P}_{gt}}
\min_{\hat{\theta}_j\in\mathcal{P}_{pred}}
d_{\mathrm{ang}}(\theta_i,\hat{\theta}_j),
\end{equation}
where $d_{\mathrm{ang}}(\cdot,\cdot)$ is the circular angular distance.

To evaluate decision-relevant arrival-direction recovery, we report Top-1 Dominant Peak Hit Rate@$\delta$ and Dominant Peak Recall@$\delta$. Hit@$\delta$ checks whether the strongest predicted peak falls within $\delta$ degrees of the strongest ground-truth dominant peak:
\begin{equation}
\mathrm{Hit@}\delta
=
\mathbf{1}
\left[
d_{\mathrm{ang}}(\theta_1,\hat{\theta}_1)
\le
\delta
\right].
\end{equation}
Recall@$\delta$ measures the fraction of ground-truth dominant peaks matched by at least one predicted peak:
\begin{equation}
\mathrm{Recall@}\delta
=
\frac{1}{|\mathcal{P}_{gt}|}
\sum_{\theta_i\in\mathcal{P}_{gt}}
\mathbf{1}
\left[
\min_{\hat{\theta}_j\in\mathcal{P}_{pred}}
d_{\mathrm{ang}}(\theta_i,\hat{\theta}_j)
\le
\delta
\right].
\end{equation}
We report $\delta=2^\circ$ and $\delta=4^\circ$, corresponding to one and two angular bins.

\subsection{Main Benchmark Results}

Table~\ref{tab:main_results} reports the main benchmark results on the full held-out test set. MS-AReg achieves the strongest overall reference performance among the compared model families, obtaining the best MAE, PSNR, CosSim, NMSE, spectral angle, and PLE among the five main benchmark models. Compared with RadioUNet, MS-AReg improves CosSim from 0.861 to 0.948, reduces MAE from 0.182 to 0.099, lowers NMSE from $-16.91$ to $-20.03$~dB, and reduces the spectral angle from $19.10^\circ$ to $10.08^\circ$. At the same time, MS-AReg runs at 0.101 ms/sample, which is about $9.2\times$ faster than RadioUNet.

A second notable comparison is against ResNet-MLP and ResNet-Conv1D, both of which have similar latency to MS-AReg but weaker prediction quality. Relative to ResNet-MLP, MS-AReg improves CosSim from 0.851 to 0.948, reduces MAE from 0.193 to 0.099, lowers PLE from $1.69^\circ$ to $1.20^\circ$, and improves NMSE from $-14.31$ to $-20.03$~dB. This shows that the accuracy gain mainly comes from better representation and training design rather than from a large increase in inference cost.

\subsection{Dominant-Direction Evaluation}

Table~\ref{tab:dominant_direction} reports the dominant-direction metrics. MS-AReg achieves the highest dominant peak recall at both $2^\circ$ and $4^\circ$, indicating that it recovers multiple decision-relevant arrival directions more reliably than the baselines. RadioUNet gives slightly higher Top-1 Hit Rate, suggesting that it can sometimes rank the strongest peak more favorably. However, RadioUNet has lower recall and weaker overall spectral-shape accuracy than MS-AReg. For beam management applications where recovering all significant arrival directions matters more than ranking a single strongest peak, recall is the more relevant metric, and MS-AReg leads consistently on both Recall@2$^\circ$ and Recall@4$^\circ$. These results show that Top-1 hit and multi-peak recall capture complementary aspects of directional reliability.

\begin{table}[t]
\centering
\caption{Dominant-direction evaluation on the full held-out test set.}
\label{tab:dominant_direction}
\scriptsize
\setlength{\tabcolsep}{2.3pt}
\begin{tabular}{lcccc}
\toprule
Method & Hit@2$^\circ$ $\uparrow$ & Hit@4$^\circ$ $\uparrow$ & Rec.@2$^\circ$ $\uparrow$ & Rec.@4$^\circ$ $\uparrow$ \\
\midrule
MS-AReg & 0.524 & 0.541 & \textbf{0.934} & \textbf{0.951} \\
RadioUNet & \textbf{0.538} & \textbf{0.558} & 0.865 & 0.893 \\
ResNet-MLP & 0.453 & 0.474 & 0.860 & 0.890 \\
ViT-Reg & 0.481 & 0.501 & 0.817 & 0.847 \\
ResNet-Conv1D & 0.457 & 0.478 & 0.818 & 0.851 \\
\bottomrule
\end{tabular}
\end{table}

\subsection{Qualitative Visualization}

Fig.~\ref{fig:qualitative} presents representative APS prediction examples on the held-out test set. The left side forms the main-comparison zone, where the ground-truth APS is overlaid with predictions from MS-AReg, RadioUNet, ResNet-MLP, and ResNet-Conv1D. The right side forms the ablation zone, where the ground truth is compared against MS-AReg, ResNet-MLP, Adv-MLP, and MS-MLP under the same overlay-style visualization. Across both zones, MS-AReg more consistently aligns dominant peaks with the ground truth and better preserves the overall spectral profile.

In the main-comparison zone, baseline methods often exhibit peak shifts, spurious responses, or over-smoothed spectra away from dominant directions, whereas MS-AReg remains closer to the ground-truth peak locations and better follows the global spectral envelope. In the ablation zone, Adv-MLP tends to improve amplitude fidelity, MS-MLP better recovers dominant peak locations, and MS-AReg achieves a stronger balance between amplitude fidelity and angular-structure preservation. Highly complex multi-peak cases remain challenging for all methods.

\subsection{Ablation Study}

Table~\ref{tab:ablation} analyzes the MS-AReg reference baseline. To isolate the roles of multi-scale encoding and training-only regularization, we compare four variants: ResNet-MLP, MS-MLP, Adv-MLP, and MS-AReg. Multi-scale encoding mainly improves spectral-shape, peak-localization, and multi-peak recovery metrics. Compared with ResNet-MLP, MS-MLP improves CosSim from 0.851 to 0.870, reduces PLE from $1.69^\circ$ to $1.31^\circ$, and improves Recall@4$^\circ$ from 0.890 to 0.913. This suggests that multi-scale fusion helps recover dominant angular structure and multiple significant directions.

Training-only regularization mainly improves amplitude-related metrics and Top-1 peak ranking. Compared with ResNet-MLP, Adv-MLP improves PSNR from 31.51 to 37.68~dB, NMSE from $-14.31$ to $-20.48$~dB, and Hit@4$^\circ$ from 0.474 to 0.546. However, its PLE becomes slightly worse and its recall remains lower than MS-AReg, indicating that amplitude fidelity, strongest-peak ranking, and multi-peak localization are not fully aligned. When both components are enabled, MS-AReg achieves the best CosSim, PLE, Recall@2$^\circ$, and Recall@4$^\circ$ among the ablation variants. These results indicate that directional localization and amplitude fidelity are complementary challenges with observable trade-offs.

\begin{table*}[t]
\centering
\caption{Ablation study of the MS-AReg reference baseline. Best results are shown in bold.}
\label{tab:ablation}
\small
\setlength{\tabcolsep}{4.2pt}
\begin{tabular}{lcccccccccc}
\toprule
Method & MS & Reg. & CosSim$\uparrow$ & PSNR$\uparrow$ & NMSE$\downarrow$ & PLE$\downarrow$ 
& Hit@2$^\circ$ $\uparrow$ & Hit@4$^\circ$ $\uparrow$ 
& Rec.@2$^\circ$ $\uparrow$ & Rec.@4$^\circ$ $\uparrow$ \\
\midrule
ResNet-MLP & $\times$ & $\times$ & 0.851 & 31.51 & -14.31 & 1.69$^\circ$ & 0.453 & 0.474 & 0.860 & 0.890 \\
MS-MLP & $\checkmark$ & $\times$ & 0.870 & 29.48 & -12.28 & 1.31$^\circ$ & 0.435 & 0.457 & 0.880 & 0.913 \\
Adv-MLP & $\times$ & $\checkmark$ & 0.853 & \textbf{37.68} & \textbf{-20.48} & 1.84$^\circ$ & \textbf{0.532} & \textbf{0.546} & 0.859 & 0.882 \\
\rowcolor{rowgray}
\textcolor{ourred}{\textbf{MS-AReg}} &
\textcolor{ourred}{$\checkmark$} &
\textcolor{ourred}{$\checkmark$} &
\textcolor{ourred}{\textbf{0.948}} &
\textcolor{ourred}{37.23} &
\textcolor{ourred}{-20.03} &
\textcolor{ourred}{\textbf{1.20$^\circ$}} &
\textcolor{ourred}{0.524} &
\textcolor{ourred}{0.541} &
\textcolor{ourred}{\textbf{0.934}} &
\textcolor{ourred}{\textbf{0.951}} \\
\bottomrule
\end{tabular}
\end{table*}

\subsection{Distributional Error Analysis}

\begin{figure}[t]
    \centering
    \includegraphics[width=0.95\columnwidth]{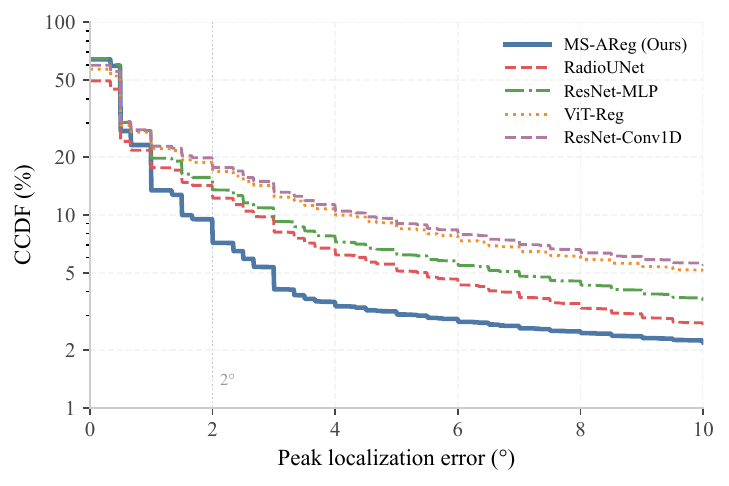}
    \caption{CCDF of peak location error (PLE) on the held-out test set. Lower is better. MS-AReg stays below the baselines across the evaluated range, indicating fewer large localization errors and distribution-wide improvement in dominant-peak prediction.}
    \label{fig:ple_ccdf}
\end{figure}

Fig.~\ref{fig:ple_ccdf} shows the PLE CCDF on the held-out test set, where MS-AReg consistently yields fewer large peak-location errors than the baselines. This indicates that the observed peak-localization improvement holds across the full evaluation distribution.

\section{Conclusion}

This paper presented Map2APS, a physically grounded benchmark for direct APS prediction from urban geometry. Map2APS defines a standardized input-output format, constructs APS labels from IRT path-level records, adopts a strict cross-map split, and evaluates models using spectral-shape, peak-localization, dominant-direction, and latency metrics. We further provided a representative benchmark suite and introduced MS-AReg as a strong reference baseline. On the full held-out test set, MS-AReg achieves the best CosSim, NMSE, spectral angle, PLE, and dominant-peak recall among the compared model families, while maintaining low inference latency. Ablation results show that multi-scale encoding mainly improves spectral-shape modeling and dominant-peak localization, whereas training-only regularization mainly improves amplitude fidelity and top-1 peak ranking. The current benchmark is restricted to equal-height 2D urban propagation and simulator-derived labels. Future work will extend Map2APS to non-uniform-height 3D environments, incorporate measured-data calibration, and include additional directional or temporal propagation descriptors.

\section*{Acknowledge}
This work was supported by the National Key Research and Development Program of China (2024YFB907500), and the National Natural Science Foundation of China (62571402).

\FloatBarrier

\bibliographystyle{IEEEtran}
\bibliography{refs}

\end{document}